\documentclass{ws-procs961x669}

\usepackage{graphicx} 
\usepackage{amsmath}
\usepackage{xcolor}
\usepackage[caption=false]{subfig}
\usepackage[T1]{fontenc}
\usepackage{float}
\usepackage{natbib}

\begin{document}

\wstoc{Disk-Outflow Symbiosis in GRMHD Simulations: Explaining Hard-State ULXs}{Mayank Pathak}

\title{Disk-Outflow Symbiosis in GRMHD Simulations: Explaining Hard-State ULXs}

\author{Mayank Pathak$^{1a}$, Banibrata Mukhopadhyay$^{2b}$}
\address{$^{1}$Joint Astronomy Programme, Department of Physics, Indian Institute of Science, Bangalore 560012, India\\
$^{2}$Department of Physics, Indian Institute of Science, Bangalore 560012, India\\
$^a$E-mail: mayankpathak@iisc.ac.in\\
$^b$E-mail: bm@iisc.ac.in\\}

\begin{abstract}
Ultraluminous X-ray sources (ULXs) have captivated researchers for decades due to their exceptionally high luminosities and unique spectral characteristics. Some of these sources defy expectations by exhibiting super-Eddington luminosities with respect to stellar mass sources even in their low-hard state. Numerical steady-state calculations suggest that ULXs in this state can be explained as highly magnetized advective accretion sources around stellar-mass black holes. To explore this further, we employ GRMHD simulations using the publicly available code, BHAC (Black Hole Accretion Code),  to model the behavior of highly magnetized advective accretion flows around a black hole. Our simulations demonstrate that such systems can indeed produce the intense luminosities observed in ULXs. Additionally, we validate that the magnetic fields required for these high emissions are of the order of $10^7$ Gauss, consistent with previous numerical steady-state findings.
\end{abstract}

\bodymatter

\section{Introduction}

Ultraluminous X-Ray sources (ULXs) are observed to have unusually high X-ray luminosities. ULXs are extremely rare, point-like, sources. Although luminosities similar to those exhibited by ULXs are not uncommon with active galactic nuclei (AGNs), ULXs are not nearly as massive as AGNs and are also non-nuclear, i.e., not located in or near galactic nuclei. These sources cannot be stars as well, as their very high luminosites, which are in the super-Eddington range, would easily tear apart a star due to radiation pressure. The Eddington luminosity ($L_{edd}$), is given by,
\begin{equation}
L_{Edd}=\frac{4\pi cGM}{\kappa_{es}},
\end{equation}
where $G$ is the gravitation constant, $M$ is the mass of the central gravitating object and $\kappa_{es}$ is the electron scattering opacity.

ULX luminosities are observed to be in the range of $3\times 10^{39} - 3\times 10^{41}$ ergs/s. Considering the dependence of $L_{Edd}$ on the mass of the central compact object, ULXs have been modelled as accreting intermediate mass black holes (IMBHs) with masses in the range of $10^2-10^4 M_{\odot}$, where $M_{\odot}$ is one solar mass. 
ULXs have also been modelled as stellar mass black holes with slim disks \cite{slim} or radiation pressure dominated super-Eddington accretion flows \cite{sup}.

Super-Eddington emissions from ULXs can also be explained by adjusting the Eddington limit. When high magnetic fields ($B\geq10^{12} G$) are present, $\kappa_{es}$ decreases \cite{ee}, lessening the impact of radiation pressure on matter. This results in a higher effective Eddington luminosity. A certain number of ULXs also show pulsations which indicate the presence of a neutron star in the system \cite{ns1}\cite{ns2}\cite{ns3}.

Although the aforementioned models can explain soft state ULX luminosities quite well, origin of hard state ULX luminosities and associated spectra is still not properly understood. A few ULX sources show very high luminosities even in their hard state. Using numerical steady state calculations, Mondal and Mukhopadhyay (MM19 hereafter) \cite{mondal} showed using their MA-AAF (magnetically arrested advective accretion flow) model that this behaviour can be well explained by considering ULXs to be highly magnetised sub-Eddington advective accretion flows around a stellar mass black hole.

In this paper, we have considered general relativistic magnetohydrodynamic (GRMHD) simulations to explore and verify the model developed in MM19. We have simulated an advective magnetized accretion flow around a stellar mass black hole with various initial conditions and have calculated outflow power from the system. This power comes out be in the range of the observed ULX luminosities. The corresponding Eulerian frame magnetic fields required to produce such high outflow power is also obtained to be in accordance with MM19.

The paper is organized as follows, In section 2, we describe our simulation setup and discuss the evolution of the system. The outflow power, magnetic field and other characteristic accretion flow profiles of our simulations are described in section 3. In section 4, we have discussed our results. We conclude in section 5.

\section{Simulation setup}
We have used the publicly available GRMHD code, BHAC (Black Hole Accretion Code) \cite{bhac} to simulate a system of an accretion disk around a Kerr black hole. We have used the Fishbone Moncrief (FM) tours setup \cite{fm} to initiate accreting matter around the black hole. The simulation uses horizon-penetrating modified Kerr-Schild (MKS) coordinates to evolve the system and data are generated in Boyer-Lindquist (BL) coordinates. We adopt geometric units, i.e., $GM_{BH}=c=1$, $r_{g}=GM_{BH}/c^2=1$ and the light crossing time, $t=GM_{BH}/c^3=1$, in our simulations.
Here, $M_{BH}$ is the mass of the black hole and $c$ is the speed of light.

The equations solved by the code are:
\begin{equation}
\begin{aligned}
    \nabla_\mu(\rho u^\mu)&=0\\
    \nabla_\mu T^{\mu\nu}&=0\\
    \nabla_{\mu}\hspace{0.01in}^*F^{\mu\nu}&=0   
\end{aligned}
\end{equation}
where, $\rho$ is the density, $u^{\mu}$ is the four-velocity, $T^{\mu\nu}$ is the stress-energy tensor and $^{*}F^{\mu\nu}$ is the dual Faraday tensor.

The black hole spin has been fixed at $a=0.9375$. We have carried out two-dimensional simulations, by exploiting the axisymmetry of the system. The computational domain extends from 1.22$r_{g}$ to 2500$r_{g}$ in the radial direction and $0$ to $2\pi$ in the azimuthal direction. The simulations have been run at a resolution of $384\times192\times1$. All simulations are evolved to $3\times10^4$ timesteps.

Accumulation of large magnetic fields in small regions in the simulation domain leads to numerical errors in GRMHD codes \cite{ress}. To avoid this, we have set the maximum value of magnetisation, i.e., $b^2/\rho$ to be 100. Whenever this limit is exceeded, matter density is injected in the coordinate frame.

\subsection{Magnetic field initiation}
The induction equation is solved to evolve the magnetic field in curved space-time. The magnetic field is initiated by defining the plasma-beta ($\beta=P_{gas}/P_{mag}$, where $P_{gas}$ is the gas pressure and $P_{mag}$ is the pressure due to the magnetic field) and initial vector potential. In our simulations, we start with a purely poloidal magnetic field with initial $\beta=100$ and have used the following vector potentials to initiate SANE (standard and normal evolution) and MAD (magnetically arrested disk) accretion systems \cite{kc}, respectively:
\begin{enumerate}
    \item $A_{\phi}=\max(\rho/\rho_0-0.2,0)$,
    \item $A_\phi=\exp(-r/r_{o})(r/r_{in})^3\sin^3\theta\max(\rho/\rho_{0}-0.01,0)$,
\end{enumerate}
where $\rho_0$ is the maximum density in the initial torus, set at $r=41r_g$, $r_{in}=20r_g$ is inner edge of the FM torus and $r_o=400 r_g$.

\begin{figure*}
\centering
 \subfloat{
\includegraphics[width=0.47\textwidth]{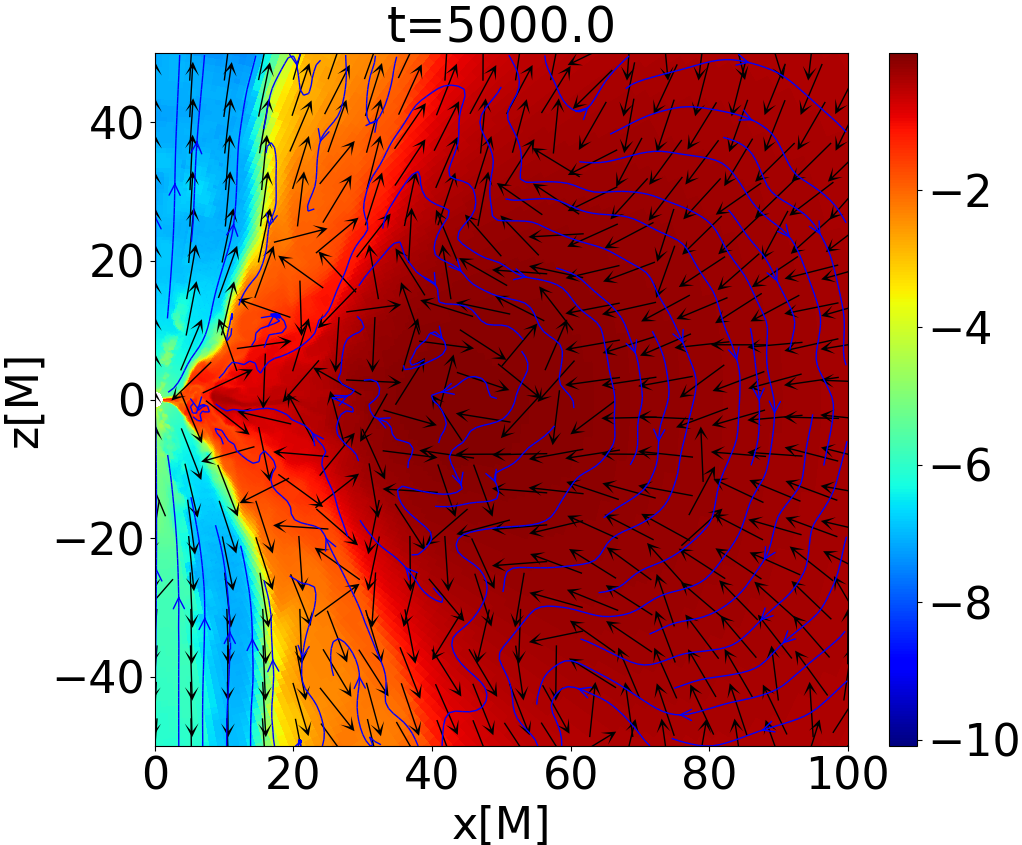}}
\subfloat{
\includegraphics[width=0.47\textwidth]{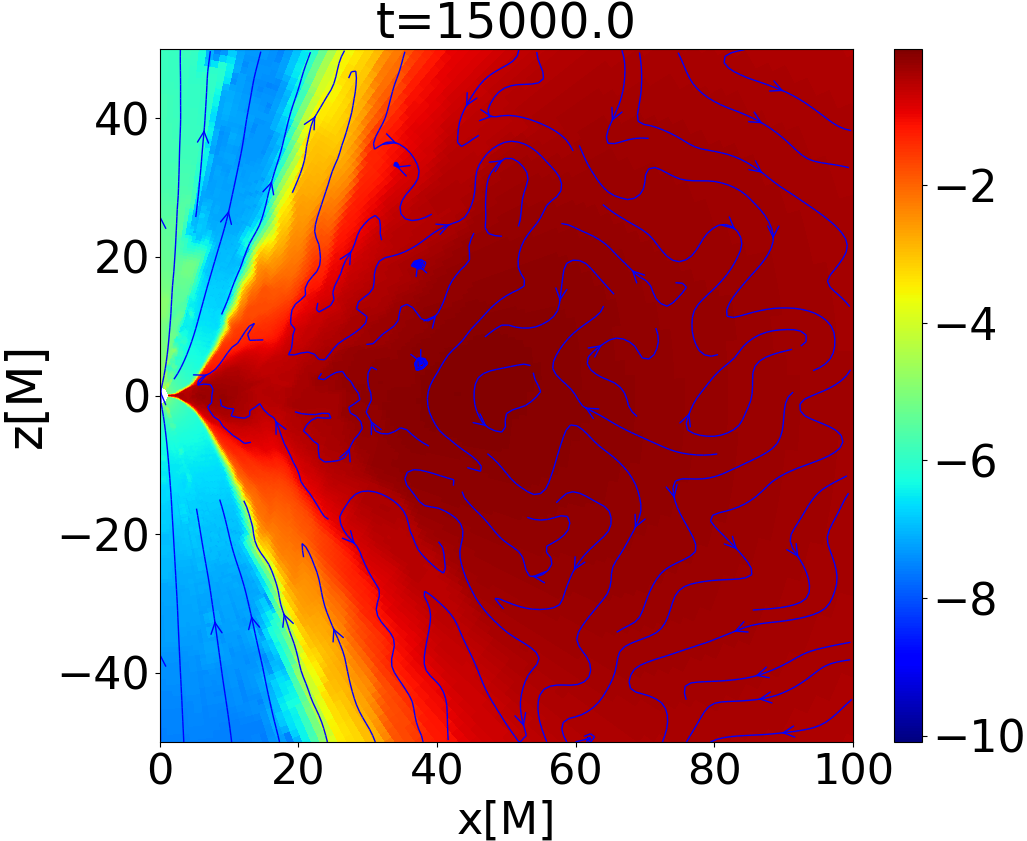}}

\subfloat{
\includegraphics[width=0.47\textwidth]{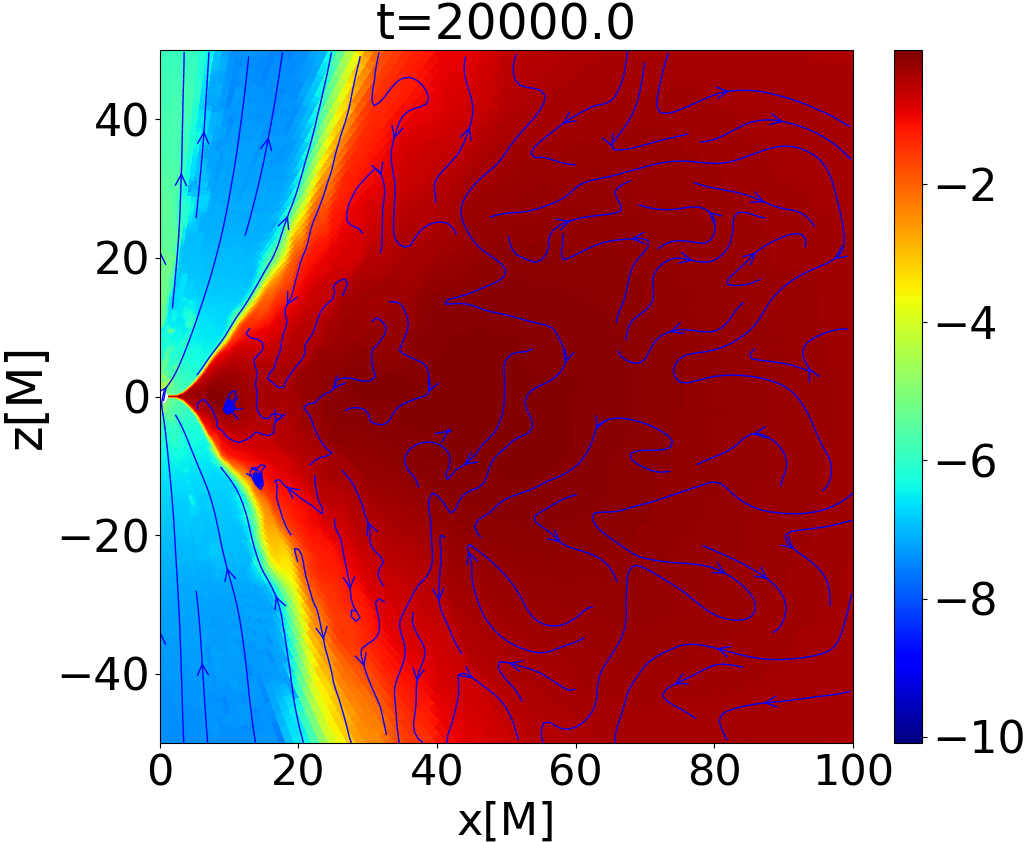}}
\subfloat{
\includegraphics[width=0.47\textwidth]{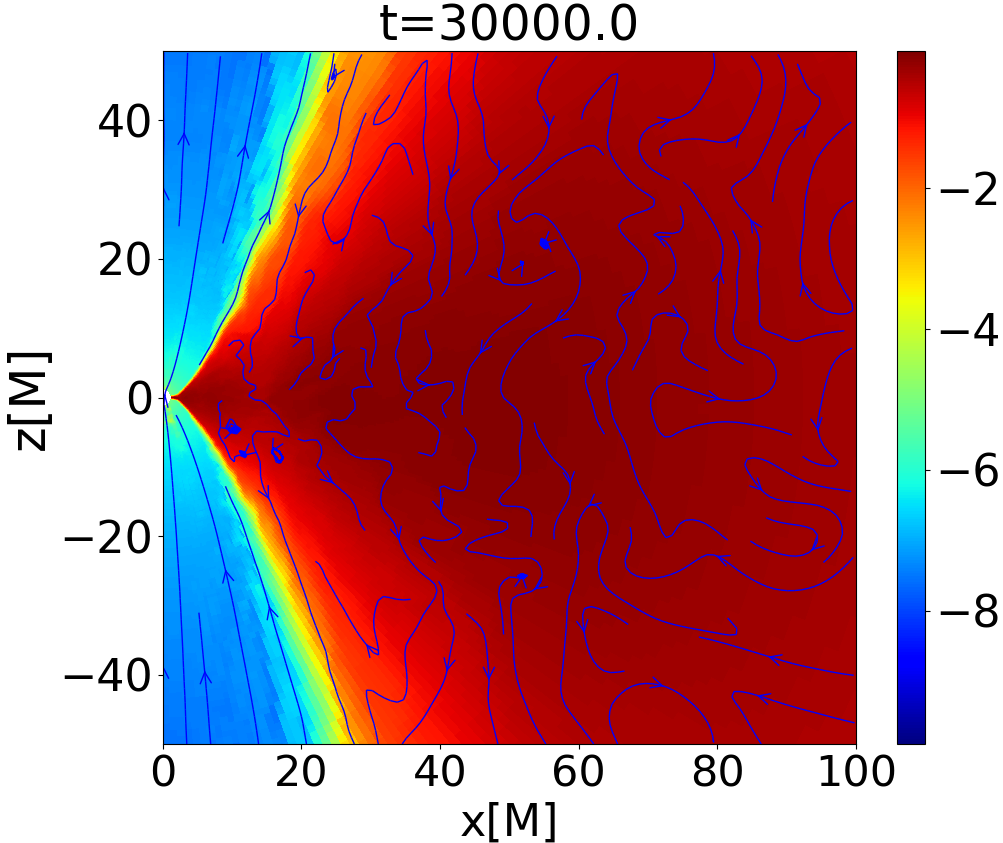}}
     \caption{Density contours for SANE with magnetic field streamlines. The top-left panel shows the velocity vectors.}
     \label{sane}
 \end{figure*}
 
\section{Results}

\begin{figure*}
\centering
\subfloat{
\includegraphics[width=0.47\textwidth]{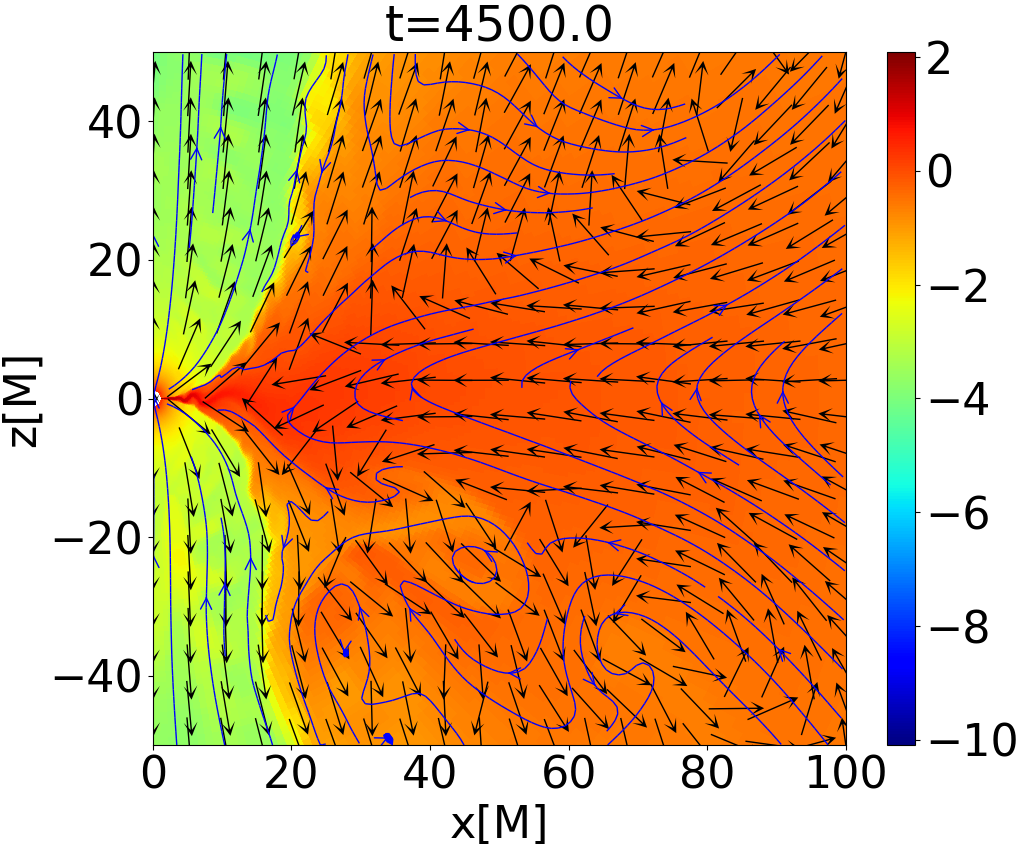}}
\subfloat{
\includegraphics[width=0.47\textwidth]{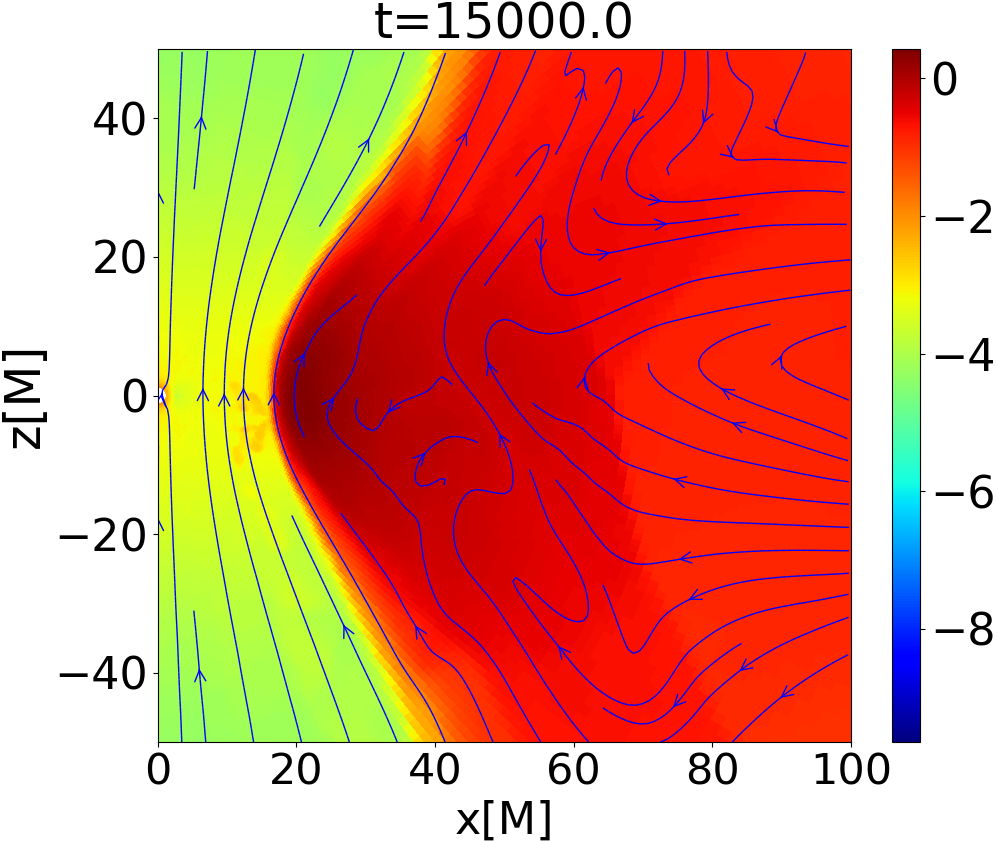}}

\subfloat{
\includegraphics[width=0.47\textwidth]{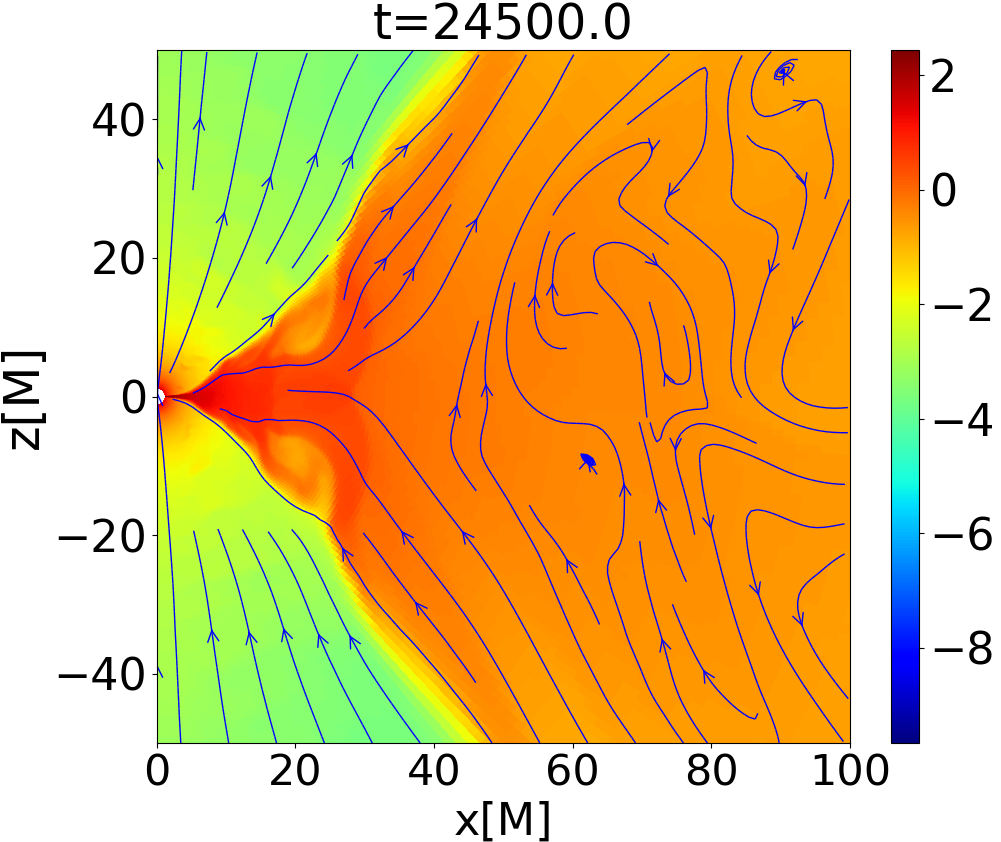}}
\subfloat{
\includegraphics[width=0.47\textwidth]{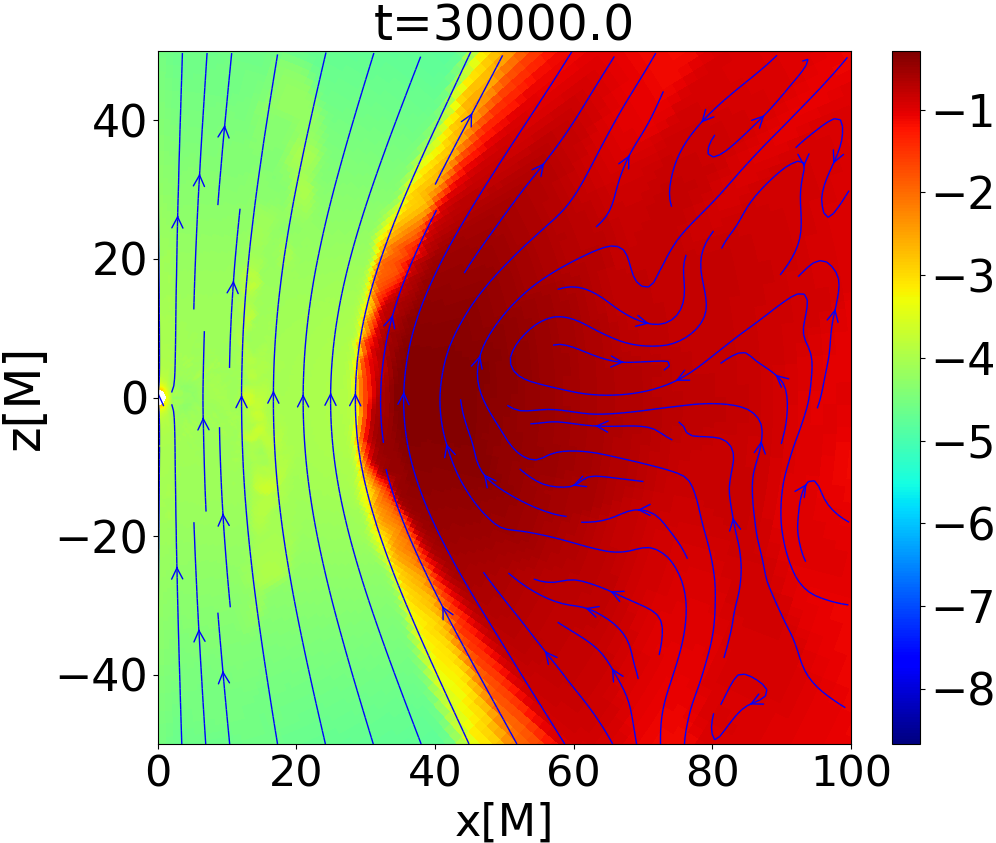}}
     \caption{Density contours for MAD with magnetic field streamlines. The top-left panel shows the velocity vectors.}
     \label{mad}
 \end{figure*}

The evolutions of flow density and magnetic fields are given in Figs. \ref{sane} and \ref{mad}. The first panel in these figures shows velocity vectors of the flow.

In SANE simulations, the accretion flow is unhindered as the simulation evolves in time. This is due to relatively weaker magnetic fields and lower magnetic flux accumulation near the black hole. At later times, the matter accretes onto the black hole through a thin funnel, due to the development of strong poloidal magnetic fields near the black hole, leading to high magnetic pressure, squeezing the accretion flow into a funnel.

In MAD simulations, on the other hand, the accretion flow is not continuous onto the black hole as evident from Fig. \ref{mad}. MAD systems reach magnetic flux saturation near the black hole in a relatively short amount of time. This leads to build up of high magnetic pressure and the formation of a magnetic barrier in the flow. As the accretion flow in our simulations is turbulent in nature, it occasionally breaks through the magnetic barrier as can be seen in the second and third panel of Fig. \ref{mad}. However, due to flux freezing, the magnetic flux accumulates again near the black hole due to the infall of matter and associated magnetic fields. This leads to the formation of the magnetic barrier again, cutting off the flow as seen in the last panel of Fig. \ref{mad}. An important point to note here is that this effect of the starting and stopping of the accretion flow is exaggerated in our two dimensional simulations as the matter has access to only one azimuthal plane for undergoing accretion. Depending on the initial conditions, the presence of additional azimuthal planes may lead to an uninterrupted accretion flow.

We use the following quantities to explore the dynamics of our accretion flow.

\begin{enumerate}
    \item Mass Accretion rate ($\Dot{M}$):
\begin{equation}
    \Dot{M}(r)=-\int\sqrt{-g}\rho u^rd\theta d\phi.
\end{equation}
    \item Inward energy flux ($\Dot{E}$):
\begin{equation}
    \Dot{E}(r)=\int \sqrt{-g}T^r_td\theta d\phi,
\end{equation}
\end{enumerate}
where $g$ is the metric determinant and $T^{\mu}_{\nu}$ is the stress energy tensor, given by
\begin{equation}
    T^{\mu}_{\nu}=\left[\rho+\tilde{P}+\tilde{P}/(\gamma+1)+b^2\right]u^{\mu}u_{\nu}+(\tilde{P}+b^2/2)\delta^{\mu}_\nu-b^{\mu}b_{\nu},
\end{equation}
where $\rho$ and $\tilde{P}$ are the density and pressure of the flow respectively, and $\gamma$ is the adiabatic constant; $u^{\mu}$ and $b^{\mu}$ are the four-velocity and four-magnetic field respectively and $b^2=b^{\mu}b_{\mu}$.

Here the signs are chosen such that positive values mean the flow of the quantity into the black hole.
The net output power is then defined as \cite{kc}:
\begin{equation}
    P(r)=(\Dot{M}(r)-\Dot{E}(r))DF=\left(-\int\sqrt{-g}\rho u^rd\theta d\phi-\int \sqrt{-g}T^r_td\theta d\phi\right)DF,\label{power}
\end{equation}
where $DF$ is the dimensional factor, used to restore the dimensions of our code unit outflow power. To infer $P(r)$ in physical units, we need to multiply it with $\Dot{M}_sc^2$, where $\Dot{M}_s$ is the scale of accretion rate, defined by $\Dot{M}_{phy}=\Dot{M}_s\Dot{M}(r')$, with $\Dot{M}_{phy}$ being the physical accretion rate we are interested in, and $r'$ is the radius chosen for defining the scale accretion rate. Since we are considering advective accretion flows, we have chosen $\Dot{M}_{phy}=0.05\Dot{M}_{edd}$, where $\Dot{M}_{edd}$ is the Eddington accretion rate, given by, $\Dot{M}_{edd}=1.39\times10^{18}(M_{BH}/M_{\odot})$ gm/s and $M_{BH}=20M_{\odot}$.

Thus, the dimensional power is given by
\begin{equation}
    P(r)=\left(\frac{\Dot{M}(r)-\Dot{E}(r)}{\Dot{M}(r')}\right)\Dot{M}_{phy}c^2 \label{dpower},
\end{equation}

\subsection{Steady flow radius}

The accretion flow in our simulation domain is subject to inflow and outflow. Over the course of the simulation, an inflow-outflow equilibrium is reached out to a certain radius \cite{rn}. This radius is the steady flow radius ($r_{eq}$). $r_{eq}$ depends on the initial magnetic vector potential and time of evolution.
To determine $r_{eq}$, we investigate the time averaged accretion rate profiles for both SANE and MAD simulations.

 \begin{figure*}
 \centering
\subfloat[SANE]{
\includegraphics[width=0.48\textwidth]{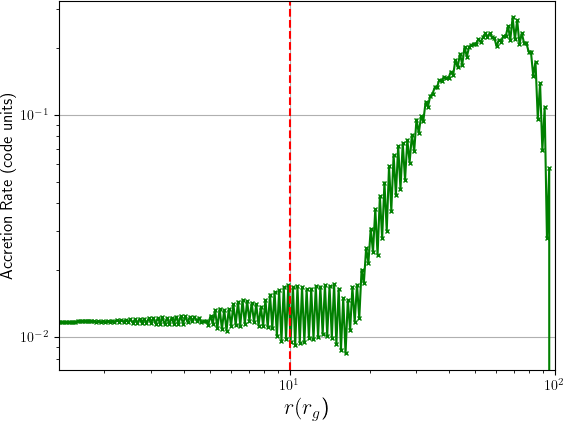}}
\subfloat[MAD]{
\includegraphics[width=0.48\textwidth]{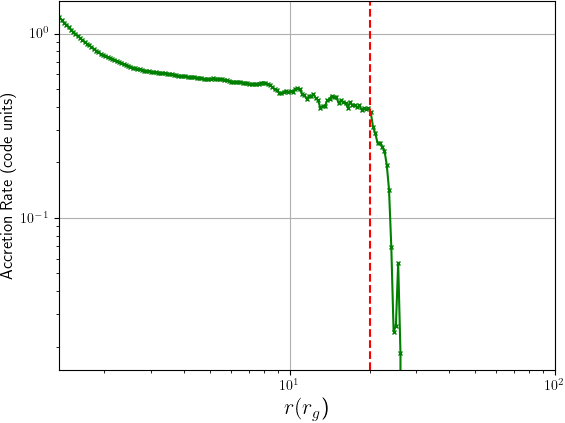}}
     \caption{Time averaged accretion rate profiles. Red dashed line indicates $r_{eq}$.}
     \label{acc}
 \end{figure*}

As evident from Fig. \ref{acc}, the radii $r_{eq}$-s for the SANE and MAD simulations are 10 $r_g$ and 20 $r_g$ respectively \cite{me}. The lower value of $r_{eq}$ for SANE indicates the slower evolution of the system compared to MAD. This is because MAD systems reach magnetic flux saturation in a smaller amount of time as compared to SANE. As both SANE and MAD systems have been evolved for the same number of timesteps, MAD has much more time to equilibrate with respect to the magnetic barrier and associated flux eruption events. This leads to higher value of $r_{eq}$ in these systems. The outflow power in our analysis has only been computed from the horizon of the black hole ($r_h$) to $r_{eq}$. In the region outside $r_{eq}$, inflow-outflow equilibrium has not been achieved by the end of our simulations and it is thus unphysical from an astrophysical standpoint.

The definition of time averaged power is ambiguous. Thus, we have considered three power definitions based on the value of $r'$:
\begin{equation}
\begin{aligned}
    P(r)&=<\Dot{M}(r)-\Dot{E}(r)>/<\Dot{M}(r)>\Dot{M}_{phy}c^2\\
    P(r)&=<\Dot{M}(r)-\Dot{E}(r)>/<\Dot{M}(r_{eq})>\Dot{M}_{phy}c^2\\
    P(r)&=<\Dot{M}(r)-\Dot{E}(r)>/<\Dot{M}(r_h)>\Dot{M}_{phy}c^2
\end{aligned}
\label{dpo}
\end{equation}

\begin{figure}
\centering
\subfloat[SANE]{
\includegraphics[width=0.83\textwidth]{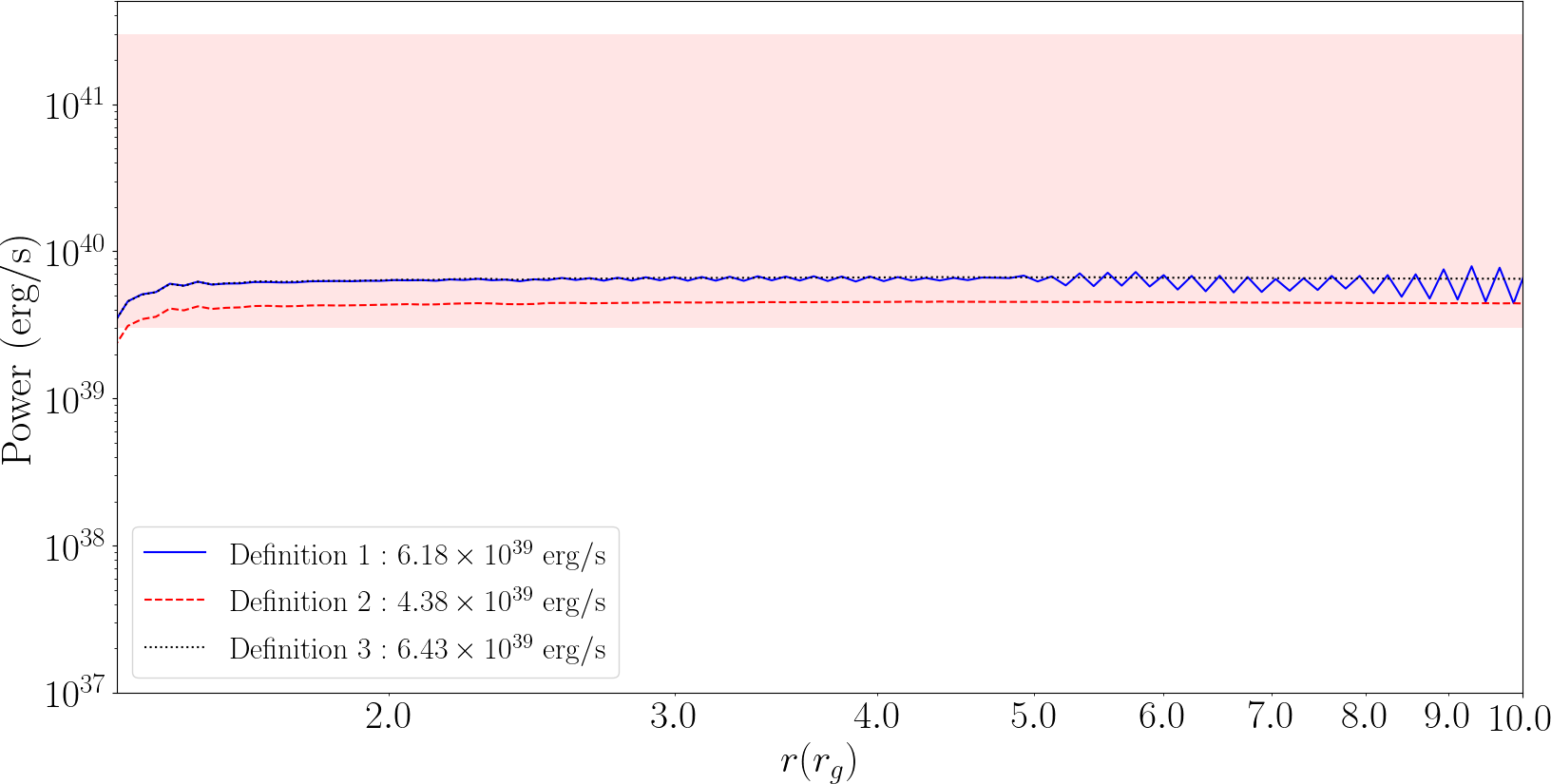}}

\subfloat[MAD]{
\includegraphics[width=0.83\textwidth]{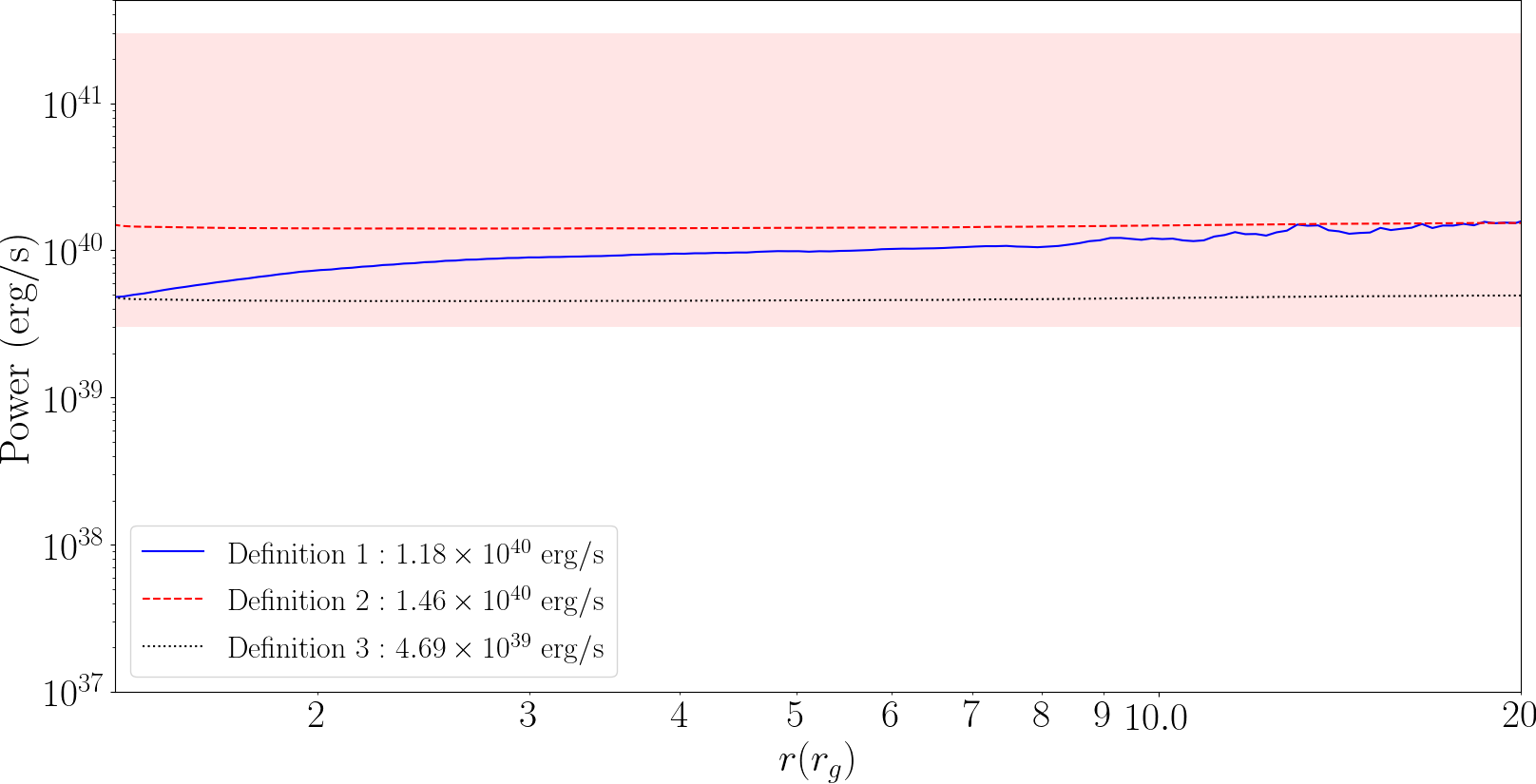}}
     \caption{Power profiles for the all definitions of time average. Shaded region indicates ULX luminosity range. Average power for the each definition of power is also mentioned next to the indicated line-type.}
     \label{po}
 \end{figure}
 
\subsection{Outflow power profiles}
In Fig. \ref{po}, we show the time averaged power profiles using the three definitions of the dimensionful power as defined in eq. (\ref{dpo}). It is clearly evident that all the power definitions lie within the ULX range for both SANE and MAD simulations. However, the average power is higher for MAD systems as compared to SANE, exhibiting the capabilities of MAD systems in producing higher outflow power. However, an important point to note here is that the outflow power has been obtained by considering the stress-energy tensor of the system. We have not considered any kind of radiative cooling in our simulations. Since the dynamics of the particles undergoing cooling will be governed by the stress energy tensor, it can be argued that the outflow power estimated here is only a precursor to ULX luminosities or the upper bound.

\subsection{Magnetic field profile}

\begin{figure}
\centering
\includegraphics[width=0.95\textwidth]{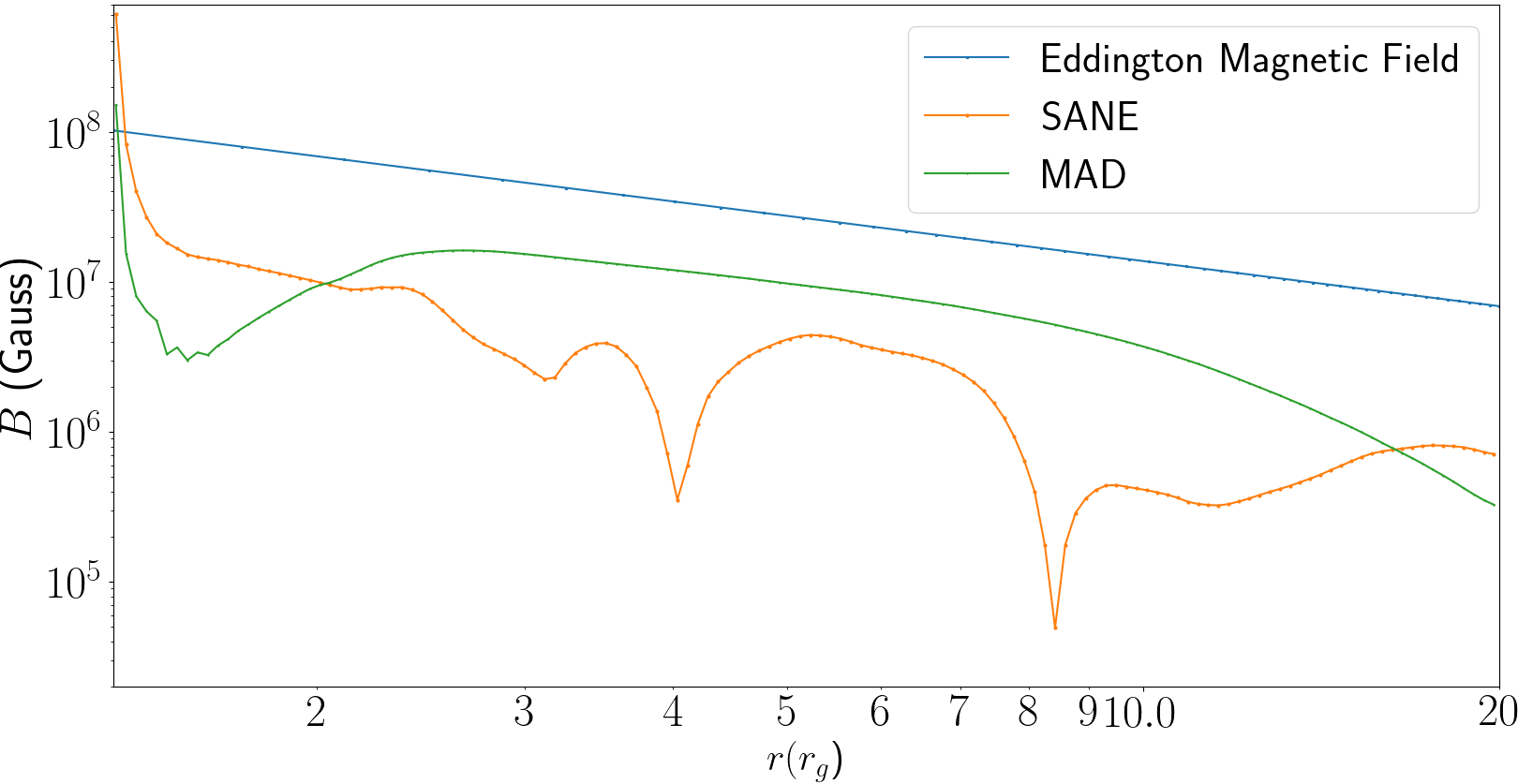}
     \caption{Magnetic field profiles for SANE and MAD simulations, overplotted with the Eddington magnetic field.}
     \label{mag}
 \end{figure}

MM19 showed that the fields required to produce ULX luminosities from advective flows are of the order of $10^7$ G. However, to achieve such high fields, they had to consider super-Eddington magnetic fields far away from the black hole. The Eddington magnetic field is defined as,
\begin{equation}
    L_{Edd}=\frac{B^2_{Edd}}{8\pi}4\pi r^2c.
\end{equation}
Fig. \ref{mag} shows the Eulerian frame magnetic field strengths obtained from our simulations, along with $B_{Edd}$.

If a system has magnetic fields of the order of $B_{Edd}$, the magnetic pressure might be high enough to destabilize it. From Fig. \ref{mag}, we can see that the magnetic fields in the SANE and MAD systems are less than $B_{Edd}$, except very close to the black hole.

The field strength obtained from our simulations is also $\sim 10^7$ G, without however any requirement for super-Eddington fields. It is also evident that although the peak magnetic field in the SANE simulations is slightly higher than the MAD simulations very close to the black hole, the higher MAD magnetic fields at other radii lead to higher luminosities in MAD systems compared to SANE. The differences in powers, whatever they be, between SANE and MAD is also reflected in their magnetic fields.

\subsection{Other flow profiles}

\begin{figure*}
\centering
\subfloat[SANE]{
\includegraphics[width=\textwidth]{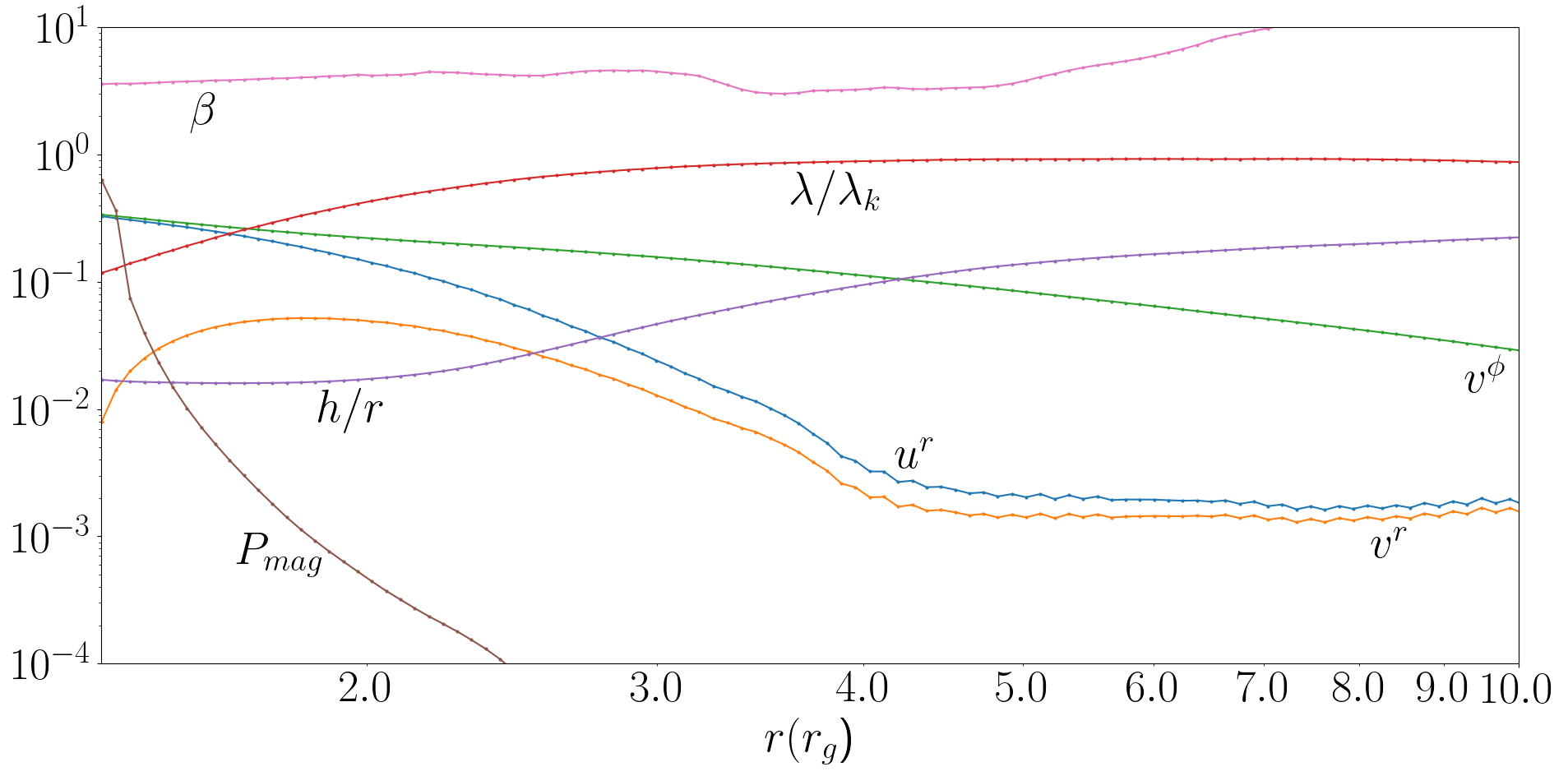}}

\subfloat[MAD]{
\includegraphics[width=\textwidth]{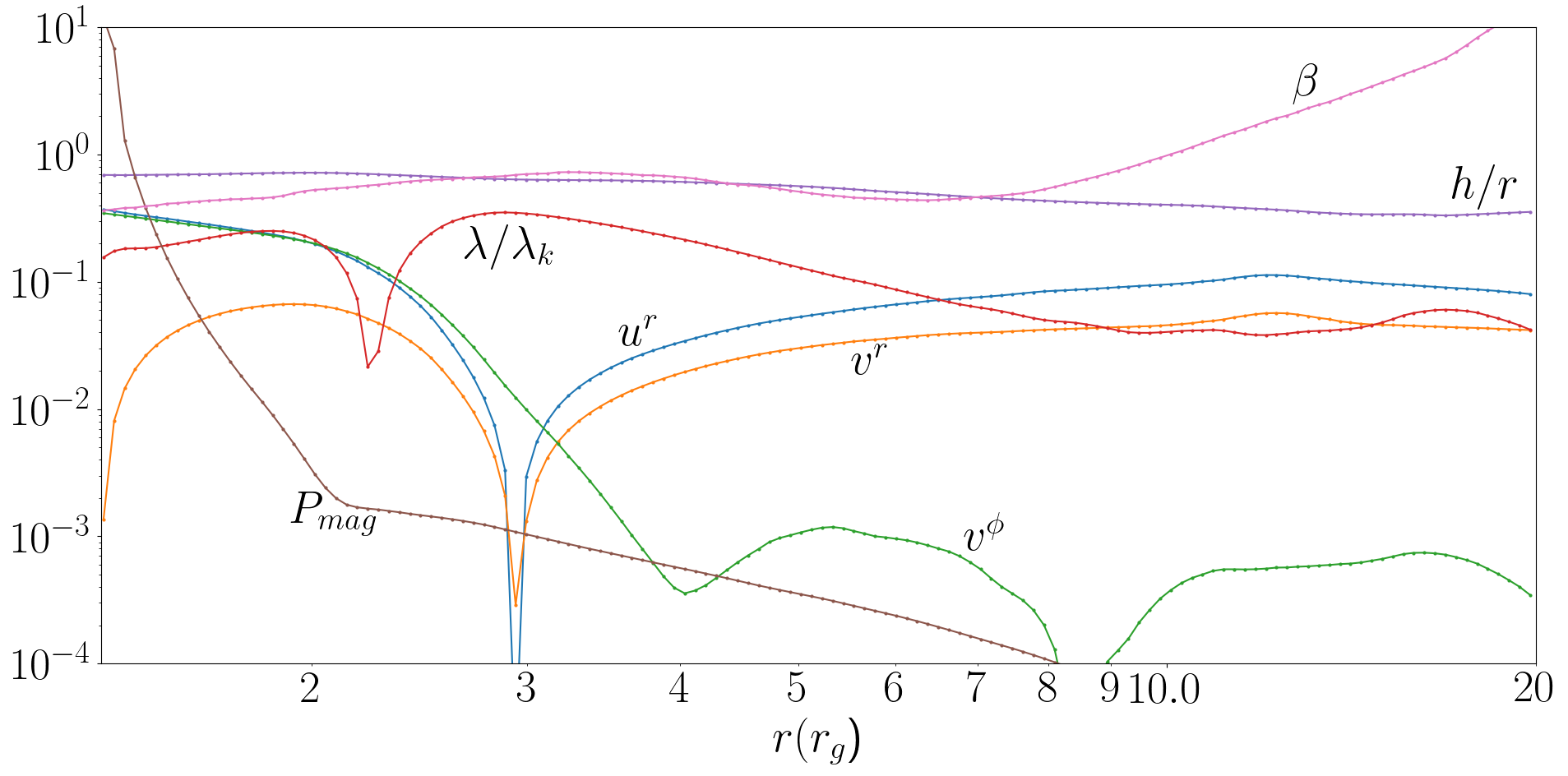}}
     \caption{Characteristic flow profiles for SANE and MAD simulations.}
     \label{vel}
 \end{figure*}

To further analyze the flow properties of the accretion system in our simulations, we have profiles of some important quantities. These profiles are given in Fig. \ref{vel}. Advective flows are characterized by a radial velocity dominated flow, i.e., the radial velocity is much more than the azimuthal velocity of the flow. This property also leads to a lower $\lambda/\lambda_k$ value as compared to thin disk accretion flows. Upon comparing the $v^r$ and $v^\phi$ profiles from Fig. \ref{vel}, it is clearly evident that MAD flows are more redial velocity dominated, except very close to the black hole. SANE flows on the other hand are $v^\phi$ dominated throughout the entire disk. Further, the $\lambda/\lambda_k$ profile also shows maximum value of around $0.4$ in the MAD case, while for SANE, this value becomes very close to $1$ above $r=3r_g$. This also shows MAD flows being advective-like while SANE flows resemble thin disk-like configuration.

The $h/r$ profiles also further corroborate this assertion. The $h/r$ in MAD simulations is quite high, close to $0.7$ throughout the entire disk. In the SANE system, however, $h/r$ is very small ($\sim 0.01$) near the black hole, and becomes $\sim 0.1$ at the edge of the inflow-outflow equilibrium region. This signifies a much thinner disk in SANE systems as compared to MAD.

Although both SANE and MAD systems are magnetically dominated, MAD systems exhibit higher magnetic dominance. This is also clear from the plasma-$\beta$ profile from Fig. \ref{vel}. The plasma-$\beta$ is around $0.3$ near the black hole in the MAD case, while it is $\sim 3$ in the SANE system. The plasma-$\beta$ profile in SANE lies consistently above the MAD plasma-$\beta$ profile throughout the entire disk. This further argues that MAD systems are more magnetically dominated than SANE. At such low plasma-$\beta$ in MAD, any thin disk-like characteristic will be completely washed over the strong magnetic fields, leading a more advection dominated flow. This further supports our assertion that MAD flows are more advective-like compared to SANE. 
 
\section{Discussion}
MAD simulations are known to show more than 100\% efficiency. This fact has been used to explain apparent more than 100\% efficient accretion flows in AGNs \cite{rn11}. In our simulations as well, we observe >100\% efficiencies in MAD systems, even though SANE has higher peak magnetic field. This is because MAD flows are very efficient in producing strong outflows. 

\begin{figure*}
\centering
\subfloat[SANE]{
\includegraphics[width=0.49\textwidth]{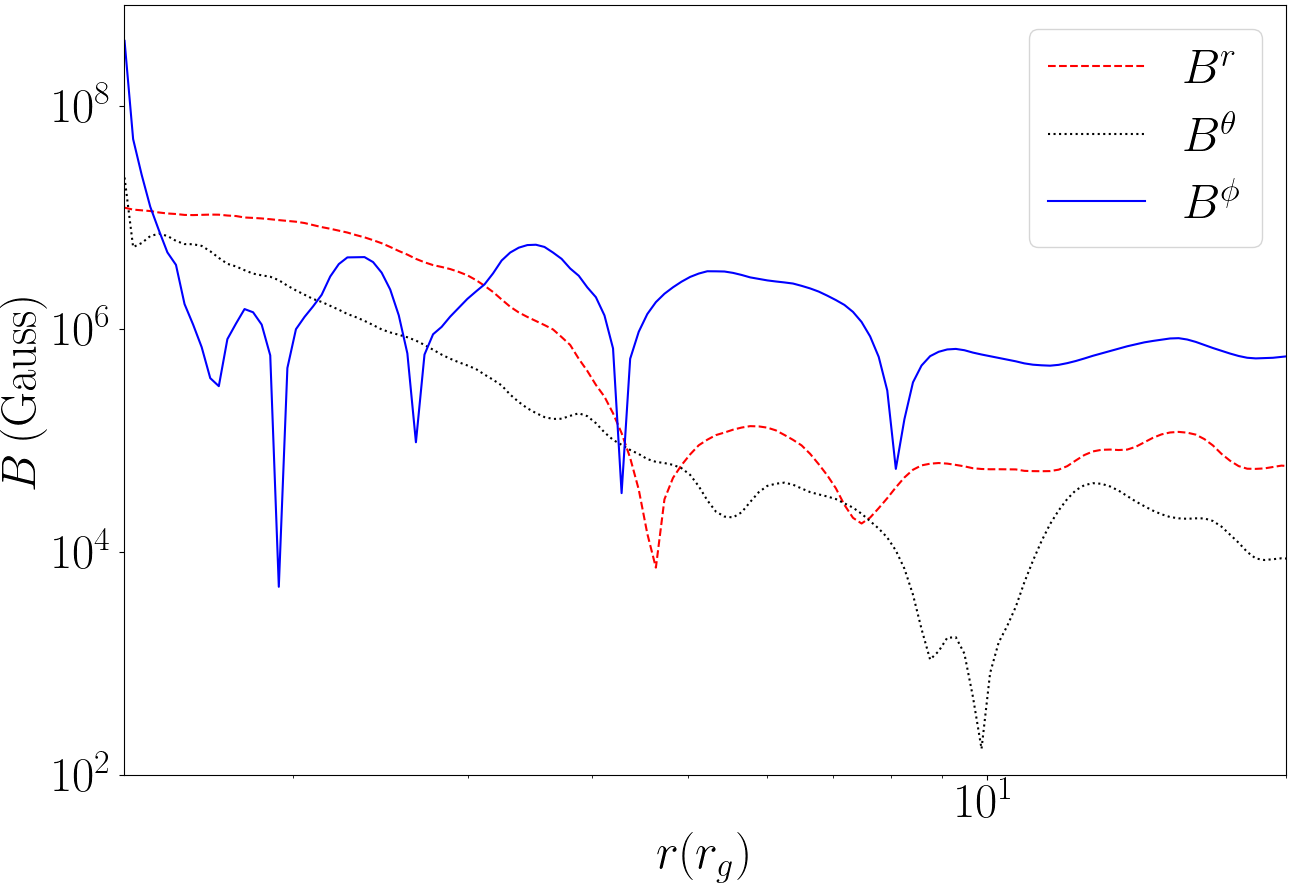}}
\subfloat[MAD]{
\includegraphics[width=0.49\textwidth]{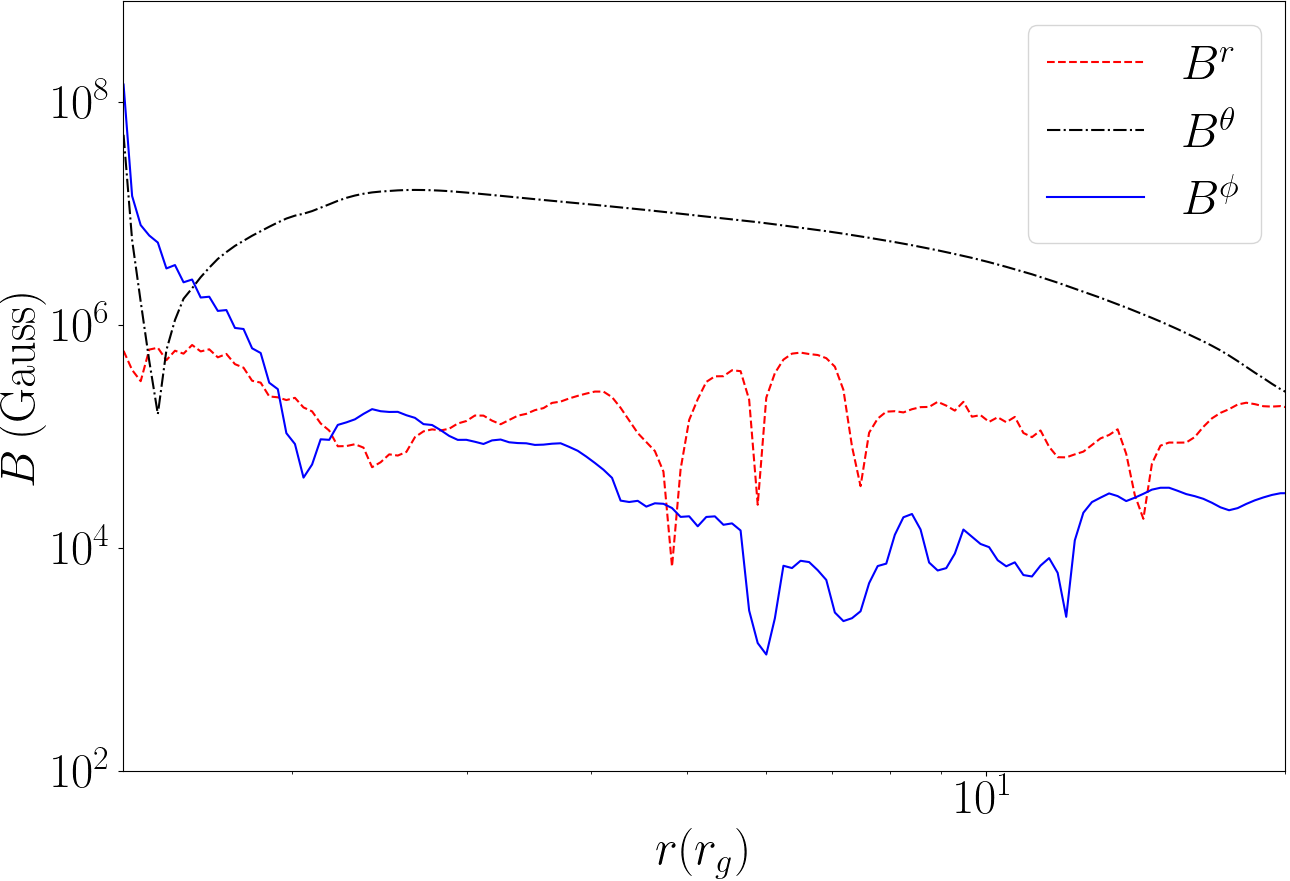}}
     \caption{Magnetic field components for SANE and MAD simulations.}
     \label{comp}
 \end{figure*}

Fig. \ref{comp} shows the Eulerian magnetic field components for SANE and MAD simulations. In SANE systems, the $B^\phi$ component dominates over the other two components. This is due to the wrapping of the field lines and hence an increase in $B^\phi$ due to the intertial frame dragging effect of the rotating black hole. All three components then reduce in magnitude on approaching away from the black hole. In MAD systems as well, $B^\phi$ is highest in magnitude close to the black hole due to the spin effect of the black hole. However, slightly away from the black hole, the $B^\theta$ component starts to dominate, indicating a poloidal field dominated system. It is this increase in $B^\theta$ at larger radii which results in sustained magnetic fields in MAD leading to higher outflow power.
This result can be attributed to the fact that magnetic fields evolve differently for MAD and SANE simulations.

\section{Conclusion}

Our simulations show that highly magnetised advective accretion flows around rotating stellar mass black holes can indeed produce high outflow power, well within the observed ULX luminosity range ($\sim 10^{40}$ erg/s). More specifically, MAD accretion flows are capable of producing higher outflow power than SANE flows. MADs also exhibit advective-like flow characteristics, which are not that prominent in SANE flows. This perfectly corroborates with the MA-AAF model.  

This shows that the peculiar properties of hard state ULXs can be explained without invoking ideas like modifying the Eddington luminosity by changing the electron scattering cross-section or considering ULXs to be intermediate mass black holes, only very few of which are currently known.

\end{document}